\documentstyle[12pt]{article}

\begin{document}

\title{Revisit of Weinberg's sum rules}
\author{Bing An Li\\
Department of Physics and Astronomy, University of Kentucky\\
Lexington, KY 40506, USA}

\maketitle

\begin{abstract}

By applying the Ward identity found by Weinberg
two new relations of the amplitude of $a_{1}\rightarrow\rho\pi$
with other physical quantities have been found.
\end{abstract}

\newpage
\newcommand{\ssp} {\partial \hspace{-.09in}/}
\newcommand{\sspp} {p \hspace{-.07in}/}
\newcommand{\pa} {\partial}
\newcommand{\sso} {\omega \hspace{-.09in}/}
\newcommand{\da} {\dagger}

\newpage
One of the most important features revealed from quantum chromodynamics
(QCD) is the chiral symmetry in the physics of ordinary hadrons.
However, chiral symmetry has been extensively applied to study
hadron physics before $QCD$ created.
Weinberg's
sum rules of $\rho$ and $a_{1}$ mesons[1] are among those pioneer works.
In ref.[1] by applying $SU(2)_{L}\times SU(2)_{R}$ and current algebra
two Ward identities of vector and axial-vector currents
have been obtained and two sum rules have been found. However, in order
to obtain the second sum rule(eq.(4) of ref.(1)) one additional
assumption has been made[1].
Using VMD and combining Weinberg's sum rules with KSFR sum rule[2],
the relationship
between the masses of $\rho$ and $a_{1}$ mesons \(m_{a_{1}}^{2}=2
m_{\rho}^{2}\) has been revealed. It predicts that \(m_{a}=1.09
GeV\). It is well known that $a_{1}$ meson has a long history
and the present mass of $a_{1}$ meson is 1.26 GeV[3].
In this letter we try
to see what we can learn if we only apply
chiral symmetry, VMD, and current algebra to
study the physics of $\rho$ and $a_{1}$ mesons. We follow
the Ward identity found by Weinberg[1]. From the formalism
of ref.[1] it can be seen that
the Weinberg's first sum rule is
the natural result of chiral symmetry, and current algebra.
Under VMD, the first sum rule
\begin{equation}
\int^{\infty}_{0}\{\rho_{V}(\mu^{2})-\rho_{A}(\mu^{2})\}\mu^{-2}
d\mu^{2}=\frac{F_{\pi}^{2}}{4}
\end{equation}
leads to
\begin{equation}
\frac{g_{\rho}^{2}}{m_{\rho}^{2}}-\frac{g_{A}^{2}}{m_{A}^{2}}
=\frac{F_{\pi}^{2}}{4}
\end{equation}
where
\[<0|V^{a}_{\mu}|\rho^{\lambda}_{b}>=\epsilon^{\lambda}_{\mu}
\delta_{ab}g_{\rho},\]
\[<0|A^{a}_{\mu}|a^{\lambda}_{b}>=\epsilon^{\lambda}_{\mu}
\delta_{ab}g_{A},\]
and $F_{\pi}$ is pion decay constant, \(F_{\pi}=186MeV\).
This sum rule(2) can be tested.
According to VMD, $g_{\rho}$ is the coupling constant between
$\rho$ and $\gamma$ and it has been determined \(g_{\rho}=0.12
GeV^{2}\) from $\Gamma(\rho\rightarrow l^{+}l^{-})$.
Using \(m_{a}=1.26 GeV\) the first sum rule predicts that
\begin{equation}
g_{A}=0.16 GeV^{2}.
\end{equation}
On the other hand, $g_{A}$ can be determined from $\Gamma
(\tau\rightarrow a_{1}\nu)$. By using VMD, we have
\begin{equation}
\Gamma(\tau\rightarrow a_{1}\nu)=\frac{G^{2}}{8\pi}cos^{2}
\theta_{c}g^{2}_{A}\frac{m^{3}_{\tau}}{m^{2}_{a}}(1-
\frac{m^{2}_{a}}{m^{2}_{\tau}})^{2}(1+2\frac{m^{2}_{a}}
{m^{2}_{\tau}}).
\end{equation}
The experimental data of the decay rate is[3] $2.14\times 10^{-13}
(1\pm 0.32)GeV$. $g_{A}$ is determined to be
$0.16(1\pm 0.16)GeV^{2}$.
This value is in good agreement with theoretical prediction(3).

By applying the VMD to the second sum rule of
ref.[1]
\begin{equation}
\int^{\infty}_{0}\{\rho_{V}(\mu^{2})-\rho_{A}(\mu^{2})\}d\mu^{2}=0
\end{equation}
the relationship
\[g_{A}=g_{\rho}\]
has been found.
Combining eq.(5) with KSFR sum rule[2] the mass relation of
$\rho$ and $a_{1}$ mesons
\begin{equation}
m^{2}_{a}=2m^{2}_{\rho}
\end{equation}
has been established. As mentioned above, the relation(6) is not
in good agreement with data.
Therefore, it is needed to reexamine the second sum rule.
{}From ref.[1] it can be seen that
for the second sum rule(5) besides chiral symmetry and
current algebra, an extra assumption
has been used. It is not
our attempt to comment on this assumption. What we want to study
is that if we do not use this assumption and
insist on chiral symmetry, VMD, and current algebra,
what else can be obtained besides Weinberg's first sum rule(1)?
In ref.[1] the following equation (eq.(11) in ref.[1])
has been established by using $SU(2)_{L}\times SU(2)_{R}$ chiral
symmetry and current algebra
\begin{equation}
{1\over 2}q_{\mu}M^{\mu\nu\lambda}(q,p)=\Delta^{\nu\lambda}_{V}(q+p)
-\Delta^{\nu\lambda}_{A}(p),
\end{equation}
where
\begin{eqnarray}
-i\epsilon_{abc}M^{\mu\nu\lambda}=\int d^{4}xd^{4}y<0|
T\{A_{a}^{\mu}(x)
A_{b}^{\nu}(y)V_{c}^{\lambda}(0)\}|0>exp\{-iqx-ipy\},\\
\delta_{ab}\Delta^{\nu\lambda}_{V}(p)=i\int d^{4}ye^{-ipy}
<0|T\{V^{\nu}_{a}(y)V^{\lambda}_{b}(0)\}|0>.
\end{eqnarray}
The first sum rule has been derived from this equation[1].
In this letter the study starts from this equation(7).
Following ref.[1] setting \(q_{\mu}=0\) in eq.(7), on the left hand of
the equation(7) only pion poles survive. In this limit the equation
(7)now reads
\begin{eqnarray}
\lefteqn{
-i\epsilon_{abc}q_{\mu}M^{\mu\nu\lambda}|_{q_{\mu}\rightarrow 0}
= } \nonumber \\
& & \frac{F_{\pi}}{2k_{0}
}\int d^{4}ye^{-ipy}k\cdot q\{\frac{\theta(y_{0})}{q_{0}+k_{0}-i
\varepsilon}e^{-i(q_{0}+k_{0})y_{0}}<\pi_{a}(\vec{k}=-\vec{q},k_{0}
)|A_{b}^{\nu}(y)V_{c}^{\lambda}(0)|0>\nonumber \\
& & +\frac{\theta(-y_{0})}{q_{0}+k_{0}-i\varepsilon}
<\pi_{a}(\vec{k}=-\vec{q},k_{0})|V_{c}^{\lambda}(0)|
A_{b}^{\nu}(y)|0>\nonumber \\
& & -\frac{\theta(y_{0})}{k_{0}-q_{0}-i\varepsilon}<0|A_{b}^{\nu}(y)
V_{c}^{\lambda}(0)|\pi_{a}(\vec{k}=\vec{q},k_{0})>\nonumber \\
& & -\frac{\theta(-y_{0})}{k_{0}-q_{0}-i\varepsilon}e^{i(k_{0}-q_{0})
y_{0}}<0|V_{c}^{\lambda}(0)A_{b}^{\nu}(y)|\pi_{a}(\vec{k}
=\vec{q},k_{0})>\}.
\end{eqnarray}
In eq.(10) the pion is on mass shell and in chiral limit \(k_{0}
=|\vec{q}|\). When \(\vec{k}=-\vec{q}\) there is
\begin{equation}
\frac{k\cdot q}{2k_{0}}\frac{1}{q_{0}+k_{0}-i\varepsilon}=
{1\over 2},
\end{equation}
and when \(\vec{k}=\vec{q}\) there is
\begin{equation}
\frac{k\cdot q}{2k_{0}}\frac{1}{k_{0}-q_{0}-i\varepsilon}=-
{1\over 2}.
\end{equation}
In the limit of $q_{\mu}\rightarrow 0$, the energy of the
pion states in eq.(10) is zero. Considering this fact and substituting
the two
equations(11,12) into eq.(10), the eq.(7) becomes
\begin{equation}
F_{\pi}\int d^{4}ye^{-ipy}<\pi_{a}|T\{A^{\nu}_{b}(y)
V^{\lambda}_{c}(0)\}|0>=-\epsilon_{abc}\{\Delta^{\nu\lambda}
_{V}(p)-\Delta^{\nu\lambda}_{A}(p)\}.
\end{equation}
Comparing with the eq.(18)
of ref.[1], there is an additional factor of two in eq.(13).
This new factor does not affect the second sum rule(5) of ref.[1]
under its assumption. However, this factor is
important for the study of this letter.
In eq.(13) $\Delta^{\nu\lambda}_{V}(p)$ has $\rho$ meson pole and
$\Delta^{\nu\lambda}_{A}(p)$ has both pion pole and $a_{1}$ meson
pole. The left hand side of the equation(13) should have these three
poles. Therefore, multiplying both sides of the equation by $p^{2}
-m^{2}$ and letting $q^{2}\rightarrow m^{2}$ the corresponding
pole term can be picked out from the equation, where $m^{2}$
is pion mass(in chiral limit pion mass is zero)
, $\rho$ meson mass, and $a_{1}$ meson mass respectively.
The poles of the left hand side of the eq.(13) can be obtained
in following way
\begin{eqnarray}
\lefteqn{ F_{\pi}\int d^{4}ye^{-ipy}<\pi_{a}|T\{A^{\nu}_{b}(y)
V^{\lambda}_{c}(0)\}
|0>=}\nonumber \\\
& & \{\theta(y_{0})e^{-iky}iF_{\pi}k_{\nu}<\pi_{a}\pi_{b}(k)|
V^{\lambda}_{c}(0)|0>+
\theta(y_{0})e^{iky}(-)iF_{\pi}k_{\nu}<\pi_{a}|V^{\lambda}_{c}(0)
|\pi_{a}> \nonumber \\
& & +\theta(y_{0})e^{-iky}g_{\rho}\varepsilon_{\lambda}^{\sigma *}
(k)<\pi_{a}|A^{\nu}_{b}(0)|\rho^{\sigma}_{c}(k)>+
\theta(y_{0})e^{iky}g_{\rho}\varepsilon^{\sigma}_{\lambda}(k)
<\pi_{a}\rho^{\sigma}_{c}(k)|A^{\nu}_{b}(0)|0>\nonumber \\
& & +\theta(y_{0})e^{-iky}g_{A}\varepsilon^{\sigma}_{\nu}(k)
<\pi_{a}a^{\sigma}_{b}(k)|V^{\lambda}_{c}(0)|0>+
\theta(y_{0})e^{iky}g_{A}\varepsilon^{\sigma *}_{\nu}(k)<\pi_{
a}|V^{\lambda}_{c}(0)|a^{\sigma}_{b}(k)>\}.
\end{eqnarray}
After using VMD, the following formulas are obtained
\begin{eqnarray}
<\pi_{a}|V^{\lambda}_{c}(0)|\pi_{b}(k)>=
i\epsilon_{abc}g_{\rho}
f_{\rho\pi\pi}\frac{k^{\lambda}}{m_{m^{2}_{\rho}}},\nonumber \\
<\pi_{a}\pi_{b}(k)|V^{\lambda}_{c}(0)|0>=
-i\epsilon_{abc}g_{\rho}
f_{\rho\pi\pi}\frac{k^{\lambda}}{m_{m^{2}_{\rho}}},
\end{eqnarray}
where $f_{\rho\pi\pi}$ is the coupling constant of the decay of $\rho
\rightarrow 2\pi$ in the limit of that the energy of one pion
is zero.
Using VMD, the other four matrix elements in eq.(14) are related to the
decay of $a_{1}\rightarrow\rho\pi$. The vertex of this decay
has been written as
\begin{equation}
\{Ag_{\mu\nu}+Bp_{\pi\mu}p_{\pi\nu}\}\epsilon_{abc}a^{a}_{\mu}\rho^{b}
_{\nu}\pi_{c},
\end{equation}
where A and B are functions of momenta of $a_{1}$,
$\rho$, and $\pi$.
In eq.(14), the energy of the state $|\pi_{a}>$ is zero,
therefore only the amplitude A contributes to these
matrix elements of eq.(14)
and the amplitude A is in the limit of \(p_{\pi}=0\).
Using eq.(16) and VMD we obtain
\begin{eqnarray}
<\pi_{a}|A^{\nu}_{b}(0)|\rho^{\sigma}_{c}(k)>=-\epsilon_{abc}g_{A}
A(m^{2}_{\rho})\varepsilon^{\sigma}(k)^{\nu}{1\over m^{2}_{\rho}-
m^{2}_{a}},\nonumber \\
<\pi_{a}\rho^{\sigma}_{c}(k)|A^{\nu}_{b}(0)|0>=-\epsilon_{abc}g_{A}
A(m^{2}_{\rho})\varepsilon^{\sigma *}(k)^{\nu}{1\over m^{2}_{\rho}-
m^{2}_{a}}.
\end{eqnarray}
In eq.(17), due to \(k^{2}=m^{2}_{\rho}\), we have
\(A(k^{2})=A(m^{2}_{\rho})\). In the same way, two other matrix
elements of eq.(14) can be written as
\begin{eqnarray}
<\pi_{a}|V^{\lambda}_{c}(0)|a^{\sigma}_{b}(k)>=-\epsilon_{abc}
g_{\rho}A(m_{a}^{2}){1\over m_{a}^{2}-m_{\rho}^{2}}\epsilon^{\sigma}(k)^
{\lambda},\nonumber \\
<\pi_{a}a^{\sigma}_{b}(k)|V^{\lambda}_{c}(0)|0>=-\epsilon_{abc}
g_{\rho}A(m_{a}^{2}){1\over m_{a}^{2}-m_{\rho}^{2}}
\epsilon^{\sigma *}(k)^{\lambda}.
\end{eqnarray}
It is necessary to point out that
due to the limit of \(p_{\pi}=0\) in eq.(13), the amplitude A is
function
of $k^{2}$ where $k$ is either the momentum of $\rho$ or $a_{1}$. It
has been found that in eq.(17) \(k^{2}=m^{2}_{\rho}\) and
\(k^{2}=m^{2}_{a}\) in eq.(18).
Substituting the six matrix elements(15,17,18)
into eq.(14), the pole terms on the
left hand side of the eq.(13) are obtained
\begin{eqnarray}
\lefteqn{ F_{\pi}\int d^{4}ye^{-ipy}<\pi_{a}|T\{A^{\nu}_{b}(y)
V^{\lambda}_{c}(0)\}
|0>|_{poles}=
ig_{\rho}f_{\rho\pi\pi}\frac{F^{2}_{\pi}}{m^{2}_{\rho}}
\frac{p^{\nu}p^{\lambda}}{p^{2}}}\nonumber \\
& & -i\epsilon_{abc}F_{\pi}g_{\rho}g_{A}\frac{1}{m^{2}_{\rho}
-m^{2}_{a}}{1\over p^{2}-m^{2}_{\rho}}(-g^{\lambda\nu}+\frac
{p^{\lambda}p^{\nu}}{m^{m^{2}_{\rho}}})\nonumber \\
& & -i\epsilon_{abc}F_{\pi}g_{\rho}g_{A}A(m^{2}_{a}){1\over
m^{2}_{a}-m^{2}_{\rho}}{1\over p^{2}-m^{2}_{a}}(-g^{\nu\lambda}+
\frac{p^{\nu}p^{\lambda}}{m^{2}_{a}}).
\end{eqnarray}
By applying the VMD to the right hand side of the equation(13),
the pole terms are obtained
\begin{eqnarray}
iF^{2}_{\pi}\frac{p^{\nu}p^{\lambda}}{p^{2}}+
ig^{2}_{\rho}{1\over p^{2}-m^{2}_{\rho}}(-g^{\nu\lambda}+
\frac{p^{\nu}p^{\lambda}}{p^{2}})
ig^{2}_{A}{1\over p^{2}-m^{2}_{a}}(-g^{\nu\lambda}+
\frac{p^{\nu}p^{\lambda}}{m^{2}_{a}}).
\end{eqnarray}
Multiply both sides of eq.(13) by $p^{2}$ and set
\(p^{2}=0 \) the pion
pole can be picked out and the following relation can be found
\begin{equation}
g_{\rho}f_{\rho\pi\pi}=m^{2}_{\rho}.
\end{equation}
Multiply eq.(13) by $p^{2}-m^{2}_{\rho}$ and set
\(p^{2}=m^{2}_{\rho} \) and
the $\rho$ pole can be picked out and the second relation is
obtained
\begin{equation}
g_{A}F_{\pi}A(m^{2}_{\rho})=-g_{\rho}(m^{2}_{a}-m^{2}_{\rho}).
\end{equation}
Multiply eq.(13) by $p^{2}-m^{2}_{a}$ and set \(p^{2}=m^{2}_{a}\)
and the $a_{1}$ pole can be picked out and the third relation
is found
\begin{equation}
g_{\rho}F_{\pi}A(m^{2}_{a})=-g_{A}(m^{2}_{a}-m^{2}_{\rho}).
\end{equation}
Let's discuss these three relations. The first
relation is well known from VMD[4].
In eqs.(22,23), if we assume that
$A(k^{2})$, where $k$ is either the momentum of $\rho$
meson or the momentum of $a_{1}$ meson, is independent of $k^{2}$,
i.e.
\begin{equation}
A(m_{\rho}^{2})=A(m_{a}^{2}),
\end{equation}
then from eqs.(22,23) it can be obtained that
\[g_{\rho}=g_{A}\]
which is just the consequence of the second sum rule of ref.[1].
However, based on chiral symmetry, VMD, and current
algebra only, we can not reach the conclusion(24). On the other hand,
from the discussion of Weinberg's first sum rule above and the current
value of $m_{a}$, the relation (24) does not have good support
theoretically and experimentally. Therefore, in general, $A(k^{2})$
depends on $k^{2}$. From eqs.(22,23), we determine that
\[A(m^{2}_{\rho})=-8.14GeV,\;\;\;A(m^{2}_{a})=-14.05GeV.\]
Therefore, $A(k^{2})$ strongly depends on $k^{2}$. In order to see
the deviation of these values from physical ones we can use the
experimental data of the ratio of d-wave to s wave[5] \(d/s=-
0.11\pm 0.02\) to determine
the B in the amplitude (16),
\[B=-1.1A. \]
Using the values of A and B, the width of $a_{1}\rightarrow\rho\pi$
calculated is higher than experimental value[3]
by one order of magnitude. Of course, the A's in eqs.(22,23) are
determined in an unphysical limit of \(p_{\pi}=0\).
Unlike the case of $\rho
\rightarrow\pi\pi$ that $f_{\rho\pi\pi}$(\(p_{\pi}=0\))
determined in eq.(21) and KSFR sum rule is very close to physical
value, the A's determined by eqs.(22,23) in the limit of \(p_{\pi}
=0\) are far away from the physical value of the amplitude.
In ref.[6] a chiral theory of mesons including pseudoscalar, vector and
axial-vector mesons has been studied and it can be considered as
a realization of chiral symmetry, VMD, and current algebra.
In this theory
Weinberg's first sum rule and relation (21) are satisfied.
It can also be seen that the amplitude
A found in this theory satisfies the
relations (22,23) in the limit of \(p_{\pi}=0\). Explanations of
why $f_{\rho\pi\pi}$ in the limit of \(p_{\pi}=0\) is very close to
the physical value and the amplitude A in the limit of \(p_{\pi}=0\)
is not,
can be found and a new mass relation of $\rho$ and
$a_{1}$ mesons has been presented.

To conclude, relations(22,23) do not lead to \(m^{2}_{a}=2m^{2}_{\rho}
\), therefore, from chiral symmetry, VMD, and current algebra alone
this mass relation can not be obtained. In the limit of
\(p_{\pi}=0\), the amplitude of
$a_{1}\rightarrow\rho\pi$ must satisfy the relations(22,23).

This research is partially supported by DE-91ER75661.

\end{document}